\documentclass[aps,prl,showpacs,showkeys,twocolumn]{revtex4-1} 
\usepackage{graphicx,amsmath,hyperref}
\usepackage{epstopdf, amssymb}

 \newcommand{\beq}{\begin{equation}}
 \newcommand{\eeq}{\end{equation}}
 \newcommand{\ba}{\begin{eqnarray}}
 \newcommand{\ea}{\end{eqnarray}}
 \newcommand{\bea}{\begin{eqnarray}}
 \newcommand{\eea}{\end{eqnarray}}
 \newcommand{\bma}{\begin{subequations}}
 \newcommand{\ema}{\end{subequations}}
 \newcommand{\bwt}{\begin{widetext}}
 \newcommand{\ewt}{\end{widetext}}
 \def\abs#1{|\,#1\,|}

\begin{document}

\title{A transversely localized light in waveguide: the analytical solution and its  potential application}

\author{Narkis M. Arslanov$^{1}$,  Ali A. Kamli$^{2}$, and Sergey A. Moiseev$^{1}$}
\email{samoi@yandex.ru}
\affiliation{$^{1}$Kazan Quantum Center, Kazan National Research Technical University, 10 K. Marx, Kazan, 420111, Russia}
\affiliation{$^{2}$Department of Physics, Jazan University, Jazan 28824 P O Box 144 Saudi Arabia}

\date{\today}


\begin{abstract} 
Investigation of light in the waveguide structures is a topical modern problem that has long historical roots.
A \emph{parallel-plate waveguide}
is a base model in these  studies and it is intensively used in  numerous investigations of nanooptics, integrated circuits and  nanoplasmonics.
In this letter we first have 
found the analytical solution for the light modes in this waveguide.
The solution provides a  clear physical picture for description of a light  
within the broadband spectral range in the  waveguide with various physical parameters. 
Potential of the analytical solutions for studies of light fields in the 
waveguides of nanooptics and nanoplasmonics has been also discussed.
\end{abstract}

\maketitle

Control and manipulations of the  spatially confined   light fields are the main objectives of photonics where various resonators  and  waveguide structures play a major role \cite{Hunsperger2009, Tanzilli2012, Haroche2006}.
In such environments, the light fields acquire new unusual properties  that  have caused a growing wave of wide interest \cite{Vesseur2013, Petersen2014, Bliokh2014, PhysRevLett.113.037401, PhysRevLett.112.167401, Aiello2015}.
These investigations  were spurred by the invention of the diverse new nanowaveguide systems in optics, photonics, plasmonics and its applications in manipulating the single photon fields   
\cite{Segal2015, PhysRevLett.112.167401, PhysRevLett.111.247401, PhysRevLett.111.046802, PhysRevLett.111.090502, Tame2013}.
Great expectations in observation  of unusual properties of confined light are also associated with using new materials, such as metamaterials, graphene etc. 
\cite{Tame2013, Veselago1967, PhysRevLett.85.3966, PhysRevLett.112.137401,  PhysRevLett.111.247401}.
Theoretical study of a  light  localized in the waveguide structures is a long-term topical problem \cite{Rayleigh1897, Sommerfeld1909, Stratton1941, Kogelnik1988, Yeh2008, Hunsperger2009, 1132809}. 
However, in spite of the numerous studies there are no analytical solutions  providing clear physical picture together with accurate mathematical description of the light fields in the various waveguides.


A parallel-plate (planar) waveguide depicted in  Fig. 1 is a basic model for numerous theoretical investigations
where a light  is localized in a small volume between the two planes. 
Such type of the waveguides is included as a key part in many  structures (e.g. Metal – insulator – metal  waveguide with bends \cite{Buckley2009}, as a part of the split-ring resonators \cite{Bahadori2014}, or as a subwavelength hole in a thick screen \cite{Zheng2009}, or as an approximation model for V-groove waveguide \cite{Bermudez-Urena2015}),  waveguide arrays, etc. (see \cite{Berini2009} and references there).
The developed nanooptics increases an interest to the theory of a light in the planar waveguides characterized by various geometry and material properties \cite{Kurokawa2007, Buckley2009, Buckley2009a, pra-2013, Lavoie2012, Ashour2013, Han2014, Zheng2009, Chen2015, Yang2015}.
Now the researchers need to use the numerical methods providing only numerical data of the calculations restricted by the concrete physical and geometrical parameters of the planar waveguides. 
Unfortunately, these studies do not give a possibility to get a deeper understanding of the analyzed processes and to predict the light properties for another values of the physical and geometrical parameters.

In this letter we develop the theoretical approach and have obtained  the transparent  analytically simple solution for a light in the parallel-plate waveguide.
For inspection of the obtained solution, we have used the comprehensive numerical calculations \cite{pra-2013} carried out in the wide frequency range. 

\begin{figure}
\centerline{\includegraphics[scale=0.5]{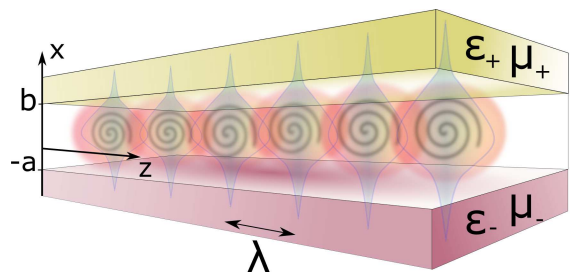}}
\vspace*{8pt}
  \caption{(Color online) 
\   The three-layer planar waveguide consists of the core layer of transverse dimension $L=a+b$, 
with  permittivity $\varepsilon_{co}(\omega)$ and magnetic permeability $\mu_{co}(\omega)$, where  $\omega=2\pi c/\lambda$, $c$ is a speed of light, $\lambda$ - the wavelength. The thicker claddings are placed at $x=b$ and $x=-a$ with $\varepsilon_{\pm}(\omega),  \mu_{\pm}(\omega)$. The light modes propagate along the waveguide axis $z$.}
  \label{fig1}
\end{figure}

In this approach we use the general properties of a transversely confined photon.
The photon wave packet confined in the spatial mode with finite cross-section  acquires the spectral dispersion in the  form of  Hamiltonian function
  $\hslash \omega=\sqrt{\left( \hslash  ck_{z} \right)^2+W_{n}}$
   corresponding to the relativistic particle with a finite mass $m_0$ \cite{DeBroglie1941, Rivlin1997, Sambles2015} propagating along $z-$direction, 
   where $W_{n}=(m_o c^2)^2=\left(\hslash ck_{n}^{\perp}\right)^2$,   $\hslash$ - Planck constant, $k_{z}$ and $k_{n}^{\perp}$ are the longitudinal and the transverse wavenumbers of the light mode $\psi_{n}$.

The  transverse confinement dramatically changes the fundamental properties of a photon and leads to the minimum frequency $\omega_{n}=m_{0} c^2/\hslash$ of n-th photon mode that is caused by the work of forces on the photon confinement. 
The crucial role of the nonzero photon mass in Bose-Einstein condensation of photons in the optical microcavity has been demonstrated in the recent experiment \cite{Klaers2010} while the important role of the nonzero photon mass in many other light effects remains an open area.  

In practice, the waveguides have various geometries,  physical properties of the used materials   \cite{Yeh2008} that   determine specific features of the photon modes $\psi_{n}$, its  rest mass $m_o$, and minimum frequency $ \omega_{n}$, respectively. 
In particularly, we have $\omega_{n}=\frac{\pi n}{L} c$ and   
$m_{0}=\frac{\pi n \hslash}{cL}$ for the planar waveguide with the transverse width $L$ and perfect metal cladding.
By using the mathematical method of B.Kacenelenbaum \cite{Kacenelenbaum:1953}, below we develop a perturbative analysis of the transverse confinement term $W_n$ and nonzero photon mass, respectively, in the planar waveguides with various geometric and physical characteristics.

Three layer planar waveguide consisting  of the thin slab and two much thicker claddings is depicted in Fig. \ref{fig1}. 
The eigen light  modes  are expressed here through the Hertz potentials \cite{Born-Wolf:1999:PO}:

\begin{equation}
\begin{array}{l}
\psi_{n}=\begin{cases}
 A_{n}e^{-k_{+}x}, &\text{$x>b$ },\\
B_{n}\sin{k_{n}^{\perp} x}+C_{n}\cos{k_{n}^{\perp} x}, &\text{$x \in (-a;b) $},\\
D_{n}e^{k_{-}x}, &\text{$x<-a$},
\end{cases}
\end{array}
\label{Equation-1_}
\end{equation}
\noindent

\noindent
where $e^{i (k_{z}z-\omega t)}$ is omitted (for notation convenience),
$k_{z}$ is the wavenumber parallel to the interface,
$k_{n}^{\perp}$ is the transverse wavenumber in the core, $k_{z}^2+{k_{n}^{\perp 2}}=k_{0}^{2}\varepsilon_{co} \mu_{co}$, 
$k_{\pm}$ are the transverse wavenumbers in the claddings  $k_{\pm}^2=k_{z}^2-k_{0}^{2}\varepsilon_{\pm} \mu_{\pm}$ 
(where indexes $"co"$ and $"\pm"$ denote the core and the claddings, see also Fig.\ref{fig1}). 
Here we restrict our attention to the transverse magnetic (TM)  modes (the transverse electric (TE)  modes can be studied in a similar way) for which the  electric and magnetic components of the modes are written as follows:

\begin{equation}
\begin{array}{l}
\vec{E}_{n}=\left( \frac{\partial^2 \psi_{n}}{\partial z \partial x},0,\frac{\partial^2 \psi_{n}}{\partial z^2}\right), 
 \vec{H}_{n}=\left(0, ik_{0}\varepsilon(x) \frac{\partial \psi_{n}}{\partial x},0\right)
\end{array}.
\label{Equation-2_}
\end{equation}

\noindent
From the boundary conditions for the confined light at the two interfaces $x=-a$ and $x=b$ 
we can get in standard way \cite{Adams1981} the dispersion relation in the form of transcendental equation:

\begin{equation}
\tan\left( k_{n}^{\perp}(a+b)-n\pi\right)=\frac{-\left( \frac{\varepsilon_{+}}{k_{+}}+\frac{\varepsilon_{-}}{k_{-}}\right) \frac{\varepsilon_{co}(\omega)}{k_{n}^{\perp}}}{\left( \frac{\varepsilon_{co}(\omega)}{k_{n}^{\perp}}\right)^2 -\frac{\varepsilon_{+}}{k_{+}} \frac{\varepsilon_{-}}{k_{-}}}
\label{Equation-3_}.
\end{equation}
 
\noindent

It is necessary to have the eigen mode dispersion relation in the form $k=k(\omega)$ that can provide principal information about the localized electromagnetic fields such as the phase and group velocities, transverse shape and propagation length of the low-losses modes(see \cite{PhysRevLett.111.046802} \cite{Berini2009} and references there). 
Unfortunately,  the transcendental equation (3)  does not have analytical solution and this is the well-known long-term problem  (see for example references \cite{Yeh2008, 1132809, Adams1981, Yuferev:2008:SIBc, Novotny1994, Lavoie2012, pra-2013}) of general analysis of light in various waveguides.   
    The existing theoretical approaches permit only the approximate  solutions of the transcendental equation in  three limiting cases:  i) near cuttoff, ii) at short-wave limit  $L \omega/c\gg1$, and iii) in the strongly asymmetrical case (see, for example \cite{Yeh2008}). 
While the nanowaveguide structures characterized by the intermediate sizes ($L \omega/c\leq 1$) are studied only by the numerical methods \cite{Adams1981, Yeh2008} (see also recent works \cite{Kurokawa2007, Buckley2009, Buckley2009a, pra-2013, Lavoie2012, Ashour2013, Han2014, Zheng2009, Chen2015, Yang2015, Bahadori2014, Bermudez-Urena2015}).

For analytical solution of $W_n$ we develop the perturbative method \cite{Kacenelenbaum:1953} for the original wave equation $\Delta\psi_{n}+k_{n}^{\perp\,2} \psi_{n}=0$  by taking into account the fundamental Hamiltonian form of the noted spectral dispersion  $\hslash\omega$.
Here, the eigen mode $\psi_{n}$ and $W_{n}$ are decomposed in a series  with respect to a small parameter  $\abs{q}<1$  as follows:

\begin{equation}
\begin{array}{l}
W_{n}=(\hslash c)^2({k_{n\, (0)}^{\perp 2}}+q {k_{n\,(1)}^{\perp 2}}+...+q^j {k^{\perp 2}_{n\,(j)}}+...),\\
\psi_{n}=\psi_{n}^{(0)}+q \psi_{n}^{(1)}+...+q^{j} \psi_{n}^{(j)}+...,\\
\end{array}
\label{Equation-4_}
\end{equation}

\noindent
where the physical meaning of $q$ is discussed below.

By inserting  \eqref{Equation-4_} in the wave equation and equating the terms of the same order to the small parameter $q$, we obtain  
$\Delta\psi_{n}^{(0)}+k_{n\, (0)}^{\perp\, 2}\psi_{n}^{(0)}=0$ and $\Delta\psi_{n}^{(1)}+k_{n\,(1)}^{\perp\, 2}\psi_{n}^{(0)}+k_{n\,(0)}^{\perp\,2}\psi_{n}^{(1)}=0$. One can (i) multiply the first equation by $\psi_{1}$, the second by $\psi_{0}$, (ii) integrate the equations over the waveguide cross section $S$ by taking into account Green's theorem and normalization $k_{n\,(0)}^{\perp 2}\int{dS\psi_{n}^{(0)\,2}}=1$,  and (iii) subsequent subtract of two integrals that will lead to: 

\begin{equation}
k_{n\,(1)}^{\perp 2}=\oint_{C}{dC\left(\psi_{n}^{(1)} \dfrac{\partial \psi_{n}^{(0)}}{\partial N}-\psi_{n}^{(0)} \dfrac{\partial \psi_{n}^{(1)}}{\partial N}  \right)}_{\bigl|C},
\label{Equation-5_}
\end{equation}

\noindent
where $C$ is the contour of the cross section $S$, $\vec{N}$ is a unit vector normal to the interface between the core and claddings.

The solution $k_{n\,(1)}^{\perp 2}$ in (5) is expressed through the boundary values of the functions $\psi_{n}^{(0)}$ and $\psi_{n}^{(1)}$ on the contour $C$. 
In order to find the function $\psi_{0}$, $\psi_{1}$ on the contour  $C$, we apply the Rytov-Leontovich boundary condition which has been successfully used for studying the light effects on  the interface between two different materials \cite{Yuferev:2008:SIBc}.
This condition  determines the relation  between the magnetic and electric fields of light mode at the boundary contour $C$ through the impedance $\zeta_{\pm}$ of claddings :

\begin{equation}
{\vec{E}_{co}}{\,}_{\bigl|C}\cong\zeta_{\pm}\left[\vec{N} \text{x} \vec{H}_{co}\right]_{\bigl|C},
\label{Equation-RL_}
\end{equation}

\noindent
where $\zeta^2_{\pm}(\omega)=\mu_{\pm}(\omega)/ \varepsilon_{\pm} (\omega)$ and it is assumed that $\zeta_{\pm}$  has the same order of smallness as the parameter $q$. It is worth noting that it is possible to use the modified  Rytov-Leontovich boundary condition for the light fields on the interfaces characterized by nonlocal spatial responses \cite{PhysRevX.4.041042}.

Substituting \eqref{Equation-2_},  \eqref{Equation-4_} in  \eqref{Equation-RL_} and equating the  zero-order terms, we find  ${\psi_{n}^{(0)}}_{\bigl|C}=0$,
this corresponds to the TM mode of the waveguide with perfect metal cladding. 
Here, we have  
 	$\psi_{n}^{(0)}=B_{n}\sin({k_{n\,(0)}^{\perp} x})+C_{n}\cos({k_{n\,(0)}^{\perp} x})$, where $B_{n}$ and $C_{n}$  	
are determined from the boundary conditions on the countour C and by normalization condition of $\psi_{n}^{(0)}$, $k_{n\,(0)}^{\perp\,2}=\left(\frac{\pi n}{b+a}\right)^2$.
By performing the same calculations with  \eqref{Equation-2_},  \eqref{Equation-4_} in \eqref{Equation-RL_} for the first order of smallness  $\zeta_{\pm}$ and by taking into account that ${\psi_{n}^{(0)}}_{\bigl|C}=0$, we get for ${\psi_{n}^{(1)}}_{\bigl|C}$:

\begin{equation}
{q\,\psi_{n}^{(1)}}_{\bigl|C}=-\frac{ik_{0}\epsilon_{co}}{k_{0}^{\perp 2}}{\zeta_{\pm}}\frac{\partial \psi_{0}}{\partial N}_{\bigl|C}.
\label{Equation-7_}
\end{equation}

\noindent
Using the fact that the impedance is independent of the transverse coordinates and substituting  $\psi_{n}^{(0)}$, and $\psi_{n}^{(1)}$ in  \eqref{Equation-5_}, then  we find  the analytical expression for the wave number in the asymmetric waveguide up to first order of the perturbation expansion:

\begin{equation}
\begin{array}{llr}
 k_{\perp}^2={\left( \frac{\pi n}{b+a}\right)}^{2}-(\zeta_{+}+\zeta_{-}) \frac{2ik_{0}\varepsilon_{d}}{b+a},\\
k_{z}^2=k_{0}^2 \varepsilon_{d}\mu_{d}-{\left( \frac{\pi n}{b+a}\right)}^{2}+(\zeta_{+}+\zeta_{-})\frac{2ik_{0}\varepsilon_{d}}{b+a}.
\end{array}
\label{Equation-8_}
\end{equation}

In case of the symmetric waveguide ($\zeta_{+}=\zeta_{-}=\zeta_{cl}$ and  $b=a$), the dispersion relation of even modes is given by:

\begin{equation}
\begin{array}{llr}
 k_{\perp}^2={\left( \frac{\pi n}{a}\right)}^{2}-\zeta_{cl} \frac{2ik_{0}\varepsilon_{d}}{a},\\
k_{z}^2=k_{0}^2 \varepsilon_{d}\mu_{d}-{\left( \frac{\pi n}{a}\right)}^{2}+\zeta_{cl} \frac{2ik_{0}\varepsilon_{d}}{a},\\
\end{array}
\label{Equation-9_}
\end{equation}

\noindent
and for the odd field modes we have:
\begin{equation}
\begin{array}{llr}
 k_{\perp}^2={\left(\frac{\pi}{2a}+ \frac{\pi n}{a}\right)}^{2}-\zeta_{cl} \frac{2ik_{0}\varepsilon_{d}}{a},\\
k_{z}^2=k_{0}^2 \varepsilon_{d}\mu_{d}-{\left(\frac{\pi}{2a}+ \frac{\pi n}{a}\right)}^{2}+\zeta_{cl} \frac{2ik_{0}\varepsilon_{d}}{a},\\
\end{array}
\label{Equation-10_}
\end{equation}
\noindent
where  $n = 0,1,2, ...$ defines a set of waveguide modes.
The analytical solutions   \eqref{Equation-8_}-\eqref{Equation-10_} of the studied problem is the main  result of this work. 
The wave numbers allow to find the field modes $\psi_{n}$ with the same accuracy.
Herein, the simple form of  \eqref{Equation-8_}-\eqref{Equation-10_} will clarify the physical analysis and provide a significant  advance in studying concrete tasks. It is worth noting a simplicity and perfect relation in \eqref{Equation-8_}-\eqref{Equation-10_} between the wave number, geometrical sizes and physical parameters of the media. At the same time the equations remain valid in a wide range of frequencies and the geometric dimensions of the waveguide.  
Below we compare \eqref{Equation-8_}-\eqref{Equation-10_}  with the numerical results obtained obtained for the waveguides with realistic physical
parameters.

The intensive recent numerical studies of the light field propagation in the planar nanowaveguide with the metamaterial cladding for the waveguides have been performed  in the works \cite{Lavoie2012, pra-2013}.  
 This analysis required quite large computing resources. 
In  case of the metamaterial-dielectric interface, the light field can be pushed out from the metamaterial into the dielectric core when dielectric permittivity and magnetic permeability of the metamaterial are both negative. 
As a result, the low-losses field modes  are excited on the dielectric/metamaterial interface  \cite{Kamli2008, PhysRevA.81.033839}.  Nanowaveguides \cite{Berini2009}  with such metamaterial claddings seem to be promising for realization of highly confined low-losses modes. 

We compare the numerical results for TM even modes \cite{Lavoie2012, pra-2013} with our analytical solution  \eqref{Equation-9_}. The metamaterial claddings were described by the Drude-like model of permittivity and permeability:

\begin{equation}
\begin{array}{l}
\varepsilon(\omega)/\varepsilon_{0}=1-\omega_{e}^{2}/(\omega(\omega+i\gamma_{e})),\\
\mu(\omega)/\mu_{0}=1+F\omega^2/(\omega_{0}^{2}-\omega(\omega+i\gamma_{m})),
\end{array}
\label{Equation-11_}
\end{equation}
\noindent

\noindent
where the electric $\gamma_{e}=2.73 \cdot 10^{13} s^{-1}$ and magnetic $\gamma_{m}=\gamma_{e}$ damping rates are much less than the carrier frequency of interest; $\gamma_{e,m}\ll\omega$,  $\omega_{e}=1.37\cdot  10^{16} s^{-1}$ is a plasma frequency of the material, $\omega_{0}=0.2 \omega_{e}$ is a binding frequency, and $F=0.5$ is a geometrical factor accounts for the magnetic oscillation strength. 

\begin{figure}
\centerline{\includegraphics[scale=0.20]{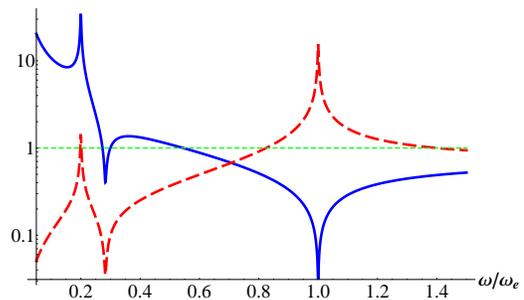}}
\caption{(Color online) Spectral behavior of the refractive index  $|n_{cl}|$ (blue line), and the impedance $|\zeta_{cl}|$  (red dashed line) of the metamaterial as functions of  $\omega/\omega_{e}$.}
\label{fig2}
\end{figure}
\begin{figure}
\centerline{\includegraphics[scale=0.40]{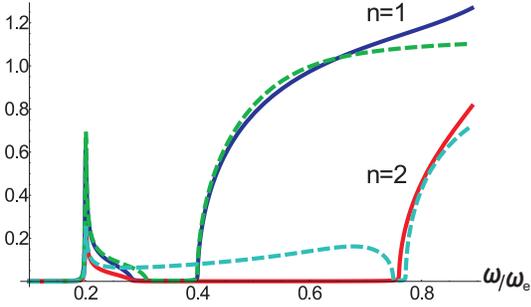}}
  \caption{(Color online) The behavior of  $Re\{k_{z}\}$ for the TM modes:  $TM_{n=1}$, $TM_{n=2}$  depending on the frequency; the analytical solutions (solid curves)  and the numerical (dashed curves) solutions are presented here for the transverse size $2a=4\pi c/\omega_{\varepsilon}\approx 275$ nm, and $\epsilon_{co}=1.3$, $\mu_{co}=1$.}
  \label{fig3}
\end{figure}

\begin{figure}

\centerline{\includegraphics[scale=0.21]{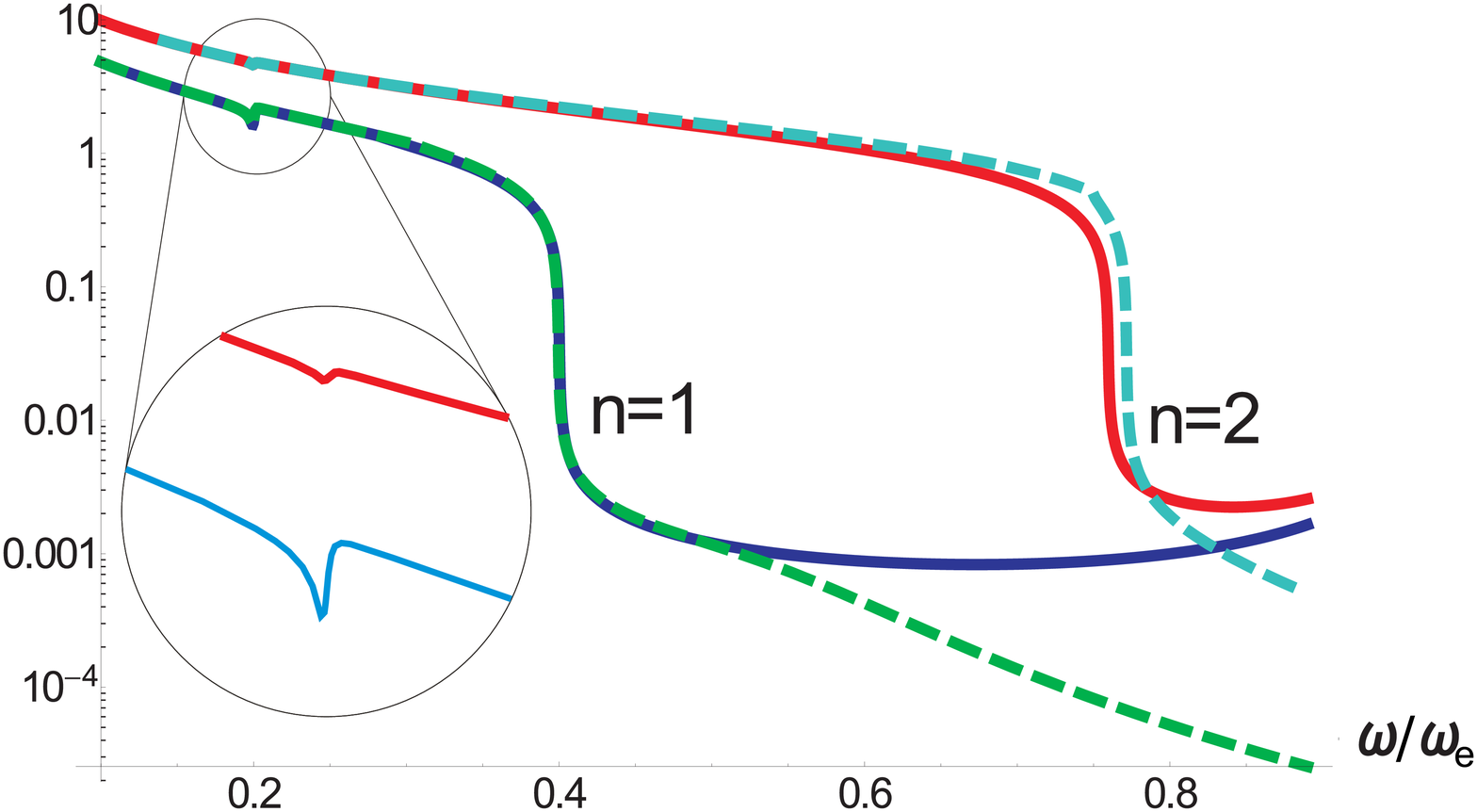}}
 \caption{(Color online) The behavior of  $Im\{k_{z}\}$ for the TM modes:  $TM_{n=1}$, $TM_{n=2}$, 
 depending on the frequency. The analytical solutions (solid curves)  and the numerical (dashed curves) solutions are given for the waveguide width $2a=4\pi c/\omega_{\varepsilon}\approx 275$ nm. The enlarged scale inset shows the sharp dips in the attenuation coefficients close to the frequency $\sim 0.2\omega/\omega_{e}$  for the both light modes (the curves demonstrate perfect coincidence between the analytical and the numerical solutions for these dips).
}
  \label{fig4}
\end{figure}

The spectral properties of the impedance $ \zeta_{cl} (\omega)$ and the refractive index $n_{cl}(\omega)$ ($n_{cl}^2(\omega)=\varepsilon (\omega)\mu(\omega)$) 
for used metamaterial cladding are presented in Fig. \ref{fig2}.
 As it is seen in Fig. \ref{fig2}, the  condition  $|\zeta_{cl}| <1$  is satisfied for the frequencies  $\omega<0.85 \omega_{e}$ (except small area around $\omega =  0.2 \omega_{e}$), so the perturbative expansion  \eqref{Equation-4_} is valid for most of the spectral range. 
It is seen in Figs. \ref{fig3},  \ref{fig4} that
each TM  mode has its spectral domain where the good match between the analytical and the numerical solutions occurs.
  In particular, for the first TM mode (n=1), we have found  the analytical solution $k_{z}^{an}$ for the longitudinal wave number coincides with the numerical one $k_{z}^{num}$ with precision $\frac{\abs{k_{z}^{an}}-\abs{k_{z}^{num}}}{\abs{k_{z}^{num}}}<10^{-2}$ for the spectral range 
  $\omega<0.25\omega_{e}$ (except for a small spectral range $\approx 0.04 \omega_{e}$ around $\omega\cong 0.2 \omega_{e}$) where the Rytov-Leontovich condition holds.
In the spectral range  $0.25 \omega_e<\omega<0.85\omega_e$, the precision of the wave number remains high  $\frac{\abs{k_{z}^{an}}-\abs{k_{z}^{num}}}{\abs{k_{z}^{num}}}<5\cdot 10^{-2}$ (i.e. in 5 times smaller, except the small area around cut-off frequency $0.4\omega_{e}$).
The solution Eq. \eqref{Equation-9_} has a large accuracy within the quite broad spectral range where the impedance is sufficiently low $\abs{\zeta_{cl}}<1$ and moreover it occurs even for the   $\abs{\zeta_{cl}}\sim 1$ provided the refractive index of cladding materials is sufficiently large  $\abs{n_{cl}}>\abs{\zeta_{cl}}$. Herein, the  higher orders of the perturbation terms in series \eqref{Equation-4_} make a negligible contribution in the analytical solution.

As it is seen in Fig.4, the difference between the analytical and the numerical solutions for attenuation coefficient is negligible within the broad spectral range of the figure. 
The good description of the approximate analytical solution is provided by the high accuracy of the Rytov-Leontovich boundary condition (6) (see, for example \cite{Guglielmi2010}) that means close to perfect metallic behavior of the waveguide cladding within the studied spectral range.    
The similar  high accuracy of the analytical description was observed in preliminary our results obtained for cylindrical waveguides with metal claddings \cite{Arslanov2005, Arslanov2007}.
Good analytical description occurs also for the attenuation coefficient  of the second TM-mode (n=2).
Since the second mode has a more complex spatial structure near the interface, the boundary conditions for the field mode should be modified in comparison with Eq.(6) in accordance to \cite{Fok1948,Sun2011} that indicates to the possible improvement of the developed analytical approach.

As seen in Fig. \ref{fig4}, the small dips in the attenuation coefficient for the both TM light modes demonstrate an emergence of the low-losses field modes (numerical and theoretical calculations are indistinguishable at the enlarged insert of the figure). 
It should be emphasized that the narrow spectral dip (with spectral accuracy $\sim 10^{-3} \omega_o$) is very well described by theoretically  along with the high accuracy of the attenuation description for the spectral range $\omega_o$. 
Our preliminary analysis of the analytical solution \eqref{Equation-9_} in this spectral range shows  the spectral dip parameters (width and depth) are highly sensitive to the waveguide transverse size, the field interference effects and the intensity distribution in the waveguide cross section. Detailed studying of the waveguides characterized by the very deep spectral dip of the light mode attenuation is very important for searching of low losses light modes and should be a subject of further investigation.

In conclusion, thus we have found the analytical solutions \eqref{Equation-8_}-\eqref{Equation-10_}  in the critical problem of theoretical  description 
of the light propagation in the nanowaveguide.  
Herein, we have developed the special perturbative approach for the light field studies that is based on using the approximate boundary conditions and general properties of Hamiltonian function of a light in such waveguides.   
The developed theory has demonstrated a robust predictive opportunity for the analytical study of transversely confined light fields within wide spectral range.

We note that the simple structure of the analytical solutions (8)-(10) and  their clear physical properties will be preserved for more complicated waveguides. 
These properties will significantly facilitate a future analysis and could provide  an advance progress for the solution a series of topical problems in nanooptics, integrated circuits and  nanoplasmonics. 
In particular, this approach can be used for understanding the properties of the waveguide excitations \cite{Bahadori2014,Kurokawa2007,Chen2015}, their interaction with localized emitters \cite{Tame2013, Chang2006, Kamli2008, Nieddu2015}  
and strong spin-orbit interaction of the light fields in the various nanowaveguides \cite{Petersen2014, Bliokh2014, Aiello2015, Cardano2015, Bliokh2015, Sukhov2015}. 
Moreover, our solutions can be valuable for detailed calculation of new metamaterials fabricated of the nanowaveguide blocks (see  \cite{Vesseur2013}), for elaboration of the waveguide light switchers \cite{Bahadori2014, Kurokawa2007, Chen2015} and for ultra-compact on-chip devices of waveguide quantum  technology  \cite{Tame2013,Manassah2012,Kamli2008,Yang2015B}. All these issues will be a subject of further analytical investigations.

The Russian Scientific Fund through the grant no. 14-12-01333 provided a financial support of N.M.A. and S.A.M.

\bibliographystyle{apsrev4-1}
\bibliography{bib}

\begin{thebibliography}{60}%
\makeatletter
\providecommand \@ifxundefined [1]{%
 \@ifx{#1\undefined}
}%
\providecommand \@ifnum [1]{%
 \ifnum #1\expandafter \@firstoftwo
 \else \expandafter \@secondoftwo
 \fi
}%
\providecommand \@ifx [1]{%
 \ifx #1\expandafter \@firstoftwo
 \else \expandafter \@secondoftwo
 \fi
}%
\providecommand \natexlab [1]{#1}%
\providecommand \enquote  [1]{``#1''}%
\providecommand \bibnamefont  [1]{#1}%
\providecommand \bibfnamefont [1]{#1}%
\providecommand \citenamefont [1]{#1}%
\providecommand \href@noop [0]{\@secondoftwo}%
\providecommand \href [0]{\begingroup \@sanitize@url \@href}%
\providecommand \@href[1]{\@@startlink{#1}\@@href}%
\providecommand \@@href[1]{\endgroup#1\@@endlink}%
\providecommand \@sanitize@url [0]{\catcode `\\12\catcode `\$12\catcode
  `\&12\catcode `\#12\catcode `\^12\catcode `\_12\catcode `\%12\relax}%
\providecommand \@@startlink[1]{}%
\providecommand \@@endlink[0]{}%
\providecommand \url  [0]{\begingroup\@sanitize@url \@url }%
\providecommand \@url [1]{\endgroup\@href {#1}{\urlprefix }}%
\providecommand \urlprefix  [0]{URL }%
\providecommand \Eprint [0]{\href }%
\providecommand \doibase [0]{http://dx.doi.org/}%
\providecommand \selectlanguage [0]{\@gobble}%
\providecommand \bibinfo  [0]{\@secondoftwo}%
\providecommand \bibfield  [0]{\@secondoftwo}%
\providecommand \translation [1]{[#1]}%
\providecommand \BibitemOpen [0]{}%
\providecommand \bibitemStop [0]{}%
\providecommand \bibitemNoStop [0]{.\EOS\space}%
\providecommand \EOS [0]{\spacefactor3000\relax}%
\providecommand \BibitemShut  [1]{\csname bibitem#1\endcsname}%
\let\auto@bib@innerbib\@empty
\bibitem [{\citenamefont {Hunsperger}(2009)}]{Hunsperger2009}%
  \BibitemOpen
  \bibfield  {author} {\bibinfo {author} {\bibfnamefont {R.}~\bibnamefont
  {Hunsperger}},\ }\href@noop {} {\emph {\bibinfo {title} {Integrated optics.
  Theory and Technology}}},\ \bibinfo {edition} {6th}\ ed.\ (\bibinfo
  {publisher} {Springer},\ \bibinfo {year} {2009})\BibitemShut {NoStop}%
\bibitem [{\citenamefont {Tanzilli}\ \emph {et~al.}(2012)\citenamefont
  {Tanzilli}, \citenamefont {Martin}, \citenamefont {Kaiser}, \citenamefont
  {De~Micheli}, \citenamefont {Alibart},\ and\ \citenamefont
  {Ostrowsky}}]{Tanzilli2012}%
  \BibitemOpen
  \bibfield  {author} {\bibinfo {author} {\bibfnamefont {S.}~\bibnamefont
  {Tanzilli}}, \bibinfo {author} {\bibfnamefont {A.}~\bibnamefont {Martin}},
  \bibinfo {author} {\bibfnamefont {F.}~\bibnamefont {Kaiser}}, \bibinfo
  {author} {\bibfnamefont {M.}~\bibnamefont {De~Micheli}}, \bibinfo {author}
  {\bibfnamefont {O.}~\bibnamefont {Alibart}}, \ and\ \bibinfo {author}
  {\bibfnamefont {D.}~\bibnamefont {Ostrowsky}},\ }\href {\doibase
  10.1002/lpor.201100010} {\bibfield  {journal} {\bibinfo  {journal} {Laser \&
  Photonics Reviews}\ }\textbf {\bibinfo {volume} {6}},\ \bibinfo {pages} {115}
  (\bibinfo {year} {2012})}\BibitemShut {NoStop}%
\bibitem [{\citenamefont {Haroche}\ and\ \citenamefont
  {Raimond}(2006)}]{Haroche2006}%
  \BibitemOpen
  \bibfield  {author} {\bibinfo {author} {\bibfnamefont {S.}~\bibnamefont
  {Haroche}}\ and\ \bibinfo {author} {\bibfnamefont {J.-M.}\ \bibnamefont
  {Raimond}},\ }\href@noop {} {\emph {\bibinfo {title} {{Exploring the Quantum:
  Atoms, Cavities, and Photons}}}}\ (\bibinfo  {publisher} {Oxford University
  Press},\ \bibinfo {address} {New York},\ \bibinfo {year} {2006})\BibitemShut
  {NoStop}%
\bibitem [{\citenamefont {Vesseur}\ \emph {et~al.}(2013)\citenamefont
  {Vesseur}, \citenamefont {Coenen}, \citenamefont {Caglayan}, \citenamefont
  {Engheta},\ and\ \citenamefont {Polman}}]{Vesseur2013}%
  \BibitemOpen
  \bibfield  {author} {\bibinfo {author} {\bibfnamefont {E.~J.~R.}\
  \bibnamefont {Vesseur}}, \bibinfo {author} {\bibfnamefont {T.}~\bibnamefont
  {Coenen}}, \bibinfo {author} {\bibfnamefont {H.}~\bibnamefont {Caglayan}},
  \bibinfo {author} {\bibfnamefont {N.}~\bibnamefont {Engheta}}, \ and\
  \bibinfo {author} {\bibfnamefont {A.}~\bibnamefont {Polman}},\ }\href
  {\doibase 10.1103/PhysRevLett.110.013902} {\bibfield  {journal} {\bibinfo
  {journal} {Phys. Rev. Lett.}\ }\textbf {\bibinfo {volume} {110}},\ \bibinfo
  {pages} {1} (\bibinfo {year} {2013})}\BibitemShut {NoStop}%
\bibitem [{\citenamefont {Petersen}\ \emph {et~al.}(2014)\citenamefont
  {Petersen}, \citenamefont {Volz},\ and\ \citenamefont
  {Rauschenbeutel}}]{Petersen2014}%
  \BibitemOpen
  \bibfield  {author} {\bibinfo {author} {\bibfnamefont {J.}~\bibnamefont
  {Petersen}}, \bibinfo {author} {\bibfnamefont {J.}~\bibnamefont {Volz}}, \
  and\ \bibinfo {author} {\bibfnamefont {A.}~\bibnamefont {Rauschenbeutel}},\
  }\href {\doibase 10.1126/science.1257671} {\bibfield  {journal} {\bibinfo
  {journal} {Science}\ }\textbf {\bibinfo {volume} {346}},\ \bibinfo {pages}
  {67} (\bibinfo {year} {2014})}\BibitemShut {NoStop}%
\bibitem [{\citenamefont {Bliokh}\ \emph {et~al.}(2014)\citenamefont {Bliokh},
  \citenamefont {Bekshaev},\ and\ \citenamefont {Nori}}]{Bliokh2014}%
  \BibitemOpen
  \bibfield  {author} {\bibinfo {author} {\bibfnamefont {K.~Y.}\ \bibnamefont
  {Bliokh}}, \bibinfo {author} {\bibfnamefont {A.~Y.}\ \bibnamefont
  {Bekshaev}}, \ and\ \bibinfo {author} {\bibfnamefont {F.}~\bibnamefont
  {Nori}},\ }\href {\doibase 10.1038/ncomms4300} {\bibfield  {journal}
  {\bibinfo  {journal} {Nat. Commun.}\ }\textbf {\bibinfo {volume} {5}}
  (\bibinfo {year} {2014}),\ 10.1038/ncomms4300}\BibitemShut {NoStop}%
\bibitem [{\citenamefont {Yang}\ \emph {et~al.}(2014)\citenamefont {Yang},
  \citenamefont {Peng}, \citenamefont {Liang}, \citenamefont {Li},\ and\
  \citenamefont {Noda}}]{PhysRevLett.113.037401}%
  \BibitemOpen
  \bibfield  {author} {\bibinfo {author} {\bibfnamefont {Y.}~\bibnamefont
  {Yang}}, \bibinfo {author} {\bibfnamefont {C.}~\bibnamefont {Peng}}, \bibinfo
  {author} {\bibfnamefont {Y.}~\bibnamefont {Liang}}, \bibinfo {author}
  {\bibfnamefont {Z.}~\bibnamefont {Li}}, \ and\ \bibinfo {author}
  {\bibfnamefont {S.}~\bibnamefont {Noda}},\ }\href {\doibase
  10.1103/PhysRevLett.113.037401} {\bibfield  {journal} {\bibinfo  {journal}
  {Phys. Rev. Lett.}\ }\textbf {\bibinfo {volume} {113}},\ \bibinfo {pages}
  {037401} (\bibinfo {year} {2014})}\BibitemShut {NoStop}%
\bibitem [{\citenamefont {Tsakmakidis}\ \emph {et~al.}(2014)\citenamefont
  {Tsakmakidis}, \citenamefont {Pickering}, \citenamefont {Hamm}, \citenamefont
  {Page},\ and\ \citenamefont {Hess}}]{PhysRevLett.112.167401}%
  \BibitemOpen
  \bibfield  {author} {\bibinfo {author} {\bibfnamefont {K.~L.}\ \bibnamefont
  {Tsakmakidis}}, \bibinfo {author} {\bibfnamefont {T.~W.}\ \bibnamefont
  {Pickering}}, \bibinfo {author} {\bibfnamefont {J.~M.}\ \bibnamefont {Hamm}},
  \bibinfo {author} {\bibfnamefont {A.~F.}\ \bibnamefont {Page}}, \ and\
  \bibinfo {author} {\bibfnamefont {O.}~\bibnamefont {Hess}},\ }\href {\doibase
  10.1103/PhysRevLett.112.167401} {\bibfield  {journal} {\bibinfo  {journal}
  {Phys. Rev. Lett.}\ }\textbf {\bibinfo {volume} {112}},\ \bibinfo {pages}
  {167401} (\bibinfo {year} {2014})}\BibitemShut {NoStop}%
\bibitem [{\citenamefont {Aiello}\ \emph {et~al.}(2015)\citenamefont {Aiello},
  \citenamefont {Banzer}, \citenamefont {Neugebauer},\ and\ \citenamefont
  {Leuchs}}]{Aiello2015}%
  \BibitemOpen
  \bibfield  {author} {\bibinfo {author} {\bibfnamefont {A.}~\bibnamefont
  {Aiello}}, \bibinfo {author} {\bibfnamefont {P.}~\bibnamefont {Banzer}},
  \bibinfo {author} {\bibfnamefont {M.}~\bibnamefont {Neugebauer}}, \ and\
  \bibinfo {author} {\bibfnamefont {G.}~\bibnamefont {Leuchs}},\ }\href
  {\doibase 10.1038/nphoton.2015.203} {\bibfield  {journal} {\bibinfo
  {journal} {Nat. Photonics}\ }\textbf {\bibinfo {volume} {9}},\ \bibinfo
  {pages} {789} (\bibinfo {year} {2015})}\BibitemShut {NoStop}%
\bibitem [{\citenamefont {Segal}\ \emph {et~al.}(2015)\citenamefont {Segal},
  \citenamefont {Keren-Zur}, \citenamefont {Hendler},\ and\ \citenamefont
  {Ellenbogen}}]{Segal2015}%
  \BibitemOpen
  \bibfield  {author} {\bibinfo {author} {\bibfnamefont {N.}~\bibnamefont
  {Segal}}, \bibinfo {author} {\bibfnamefont {S.}~\bibnamefont {Keren-Zur}},
  \bibinfo {author} {\bibfnamefont {N.}~\bibnamefont {Hendler}}, \ and\
  \bibinfo {author} {\bibfnamefont {T.}~\bibnamefont {Ellenbogen}},\ }\href
  {\doibase 10.1038/nphoton.2015.17} {\bibfield  {journal} {\bibinfo  {journal}
  {Nature Photonics}\ }\textbf {\bibinfo {volume} {9}},\ \bibinfo {pages}
  {180–184} (\bibinfo {year} {2015})}\BibitemShut {NoStop}%
\bibitem [{\citenamefont {Gullans}\ \emph {et~al.}(2013)\citenamefont
  {Gullans}, \citenamefont {Chang}, \citenamefont {Koppens}, \citenamefont
  {de~Abajo},\ and\ \citenamefont {Lukin}}]{PhysRevLett.111.247401}%
  \BibitemOpen
  \bibfield  {author} {\bibinfo {author} {\bibfnamefont {M.}~\bibnamefont
  {Gullans}}, \bibinfo {author} {\bibfnamefont {D.~E.}\ \bibnamefont {Chang}},
  \bibinfo {author} {\bibfnamefont {F.~H.~L.}\ \bibnamefont {Koppens}},
  \bibinfo {author} {\bibfnamefont {F.~J.~G.}\ \bibnamefont {de~Abajo}}, \ and\
  \bibinfo {author} {\bibfnamefont {M.~D.}\ \bibnamefont {Lukin}},\ }\href
  {\doibase 10.1103/PhysRevLett.111.247401} {\bibfield  {journal} {\bibinfo
  {journal} {Phys. Rev. Lett.}\ }\textbf {\bibinfo {volume} {111}},\ \bibinfo
  {pages} {247401} (\bibinfo {year} {2013})}\BibitemShut {NoStop}%
\bibitem [{\citenamefont {Lei\ss{}ner}\ \emph {et~al.}(2013)\citenamefont
  {Lei\ss{}ner}, \citenamefont {Lemke}, \citenamefont {Fiutowski},
  \citenamefont {Radke}, \citenamefont {Klick}, \citenamefont {Tavares},
  \citenamefont {Kjelstrup-Hansen}, \citenamefont {Rubahn},\ and\ \citenamefont
  {Bauer}}]{PhysRevLett.111.046802}%
  \BibitemOpen
  \bibfield  {author} {\bibinfo {author} {\bibfnamefont {T.}~\bibnamefont
  {Lei\ss{}ner}}, \bibinfo {author} {\bibfnamefont {C.}~\bibnamefont {Lemke}},
  \bibinfo {author} {\bibfnamefont {J.}~\bibnamefont {Fiutowski}}, \bibinfo
  {author} {\bibfnamefont {J.~W.}\ \bibnamefont {Radke}}, \bibinfo {author}
  {\bibfnamefont {A.}~\bibnamefont {Klick}}, \bibinfo {author} {\bibfnamefont
  {L.}~\bibnamefont {Tavares}}, \bibinfo {author} {\bibfnamefont
  {J.}~\bibnamefont {Kjelstrup-Hansen}}, \bibinfo {author} {\bibfnamefont
  {H.-G.}\ \bibnamefont {Rubahn}}, \ and\ \bibinfo {author} {\bibfnamefont
  {M.}~\bibnamefont {Bauer}},\ }\href {\doibase 10.1103/PhysRevLett.111.046802}
  {\bibfield  {journal} {\bibinfo  {journal} {Phys. Rev. Lett.}\ }\textbf
  {\bibinfo {volume} {111}},\ \bibinfo {pages} {046802} (\bibinfo {year}
  {2013})}\BibitemShut {NoStop}%
\bibitem [{\citenamefont {Zheng}\ \emph {et~al.}(2013)\citenamefont {Zheng},
  \citenamefont {Gauthier},\ and\ \citenamefont
  {Baranger}}]{PhysRevLett.111.090502}%
  \BibitemOpen
  \bibfield  {author} {\bibinfo {author} {\bibfnamefont {H.}~\bibnamefont
  {Zheng}}, \bibinfo {author} {\bibfnamefont {D.~J.}\ \bibnamefont {Gauthier}},
  \ and\ \bibinfo {author} {\bibfnamefont {H.~U.}\ \bibnamefont {Baranger}},\
  }\href {\doibase 10.1103/PhysRevLett.111.090502} {\bibfield  {journal}
  {\bibinfo  {journal} {Phys. Rev. Lett.}\ }\textbf {\bibinfo {volume} {111}},\
  \bibinfo {pages} {090502} (\bibinfo {year} {2013})}\BibitemShut {NoStop}%
\bibitem [{\citenamefont {Tame}\ \emph {et~al.}(2013)\citenamefont {Tame},
  \citenamefont {McEnery}, \citenamefont {Ozdemir}, \citenamefont {Lee},
  \citenamefont {Maier},\ and\ \citenamefont {Kim}}]{Tame2013}%
  \BibitemOpen
  \bibfield  {author} {\bibinfo {author} {\bibfnamefont {M.~S.}\ \bibnamefont
  {Tame}}, \bibinfo {author} {\bibfnamefont {K.~R.}\ \bibnamefont {McEnery}},
  \bibinfo {author} {\bibfnamefont {S.~K.}\ \bibnamefont {Ozdemir}}, \bibinfo
  {author} {\bibfnamefont {J.}~\bibnamefont {Lee}}, \bibinfo {author}
  {\bibfnamefont {S.~A.}\ \bibnamefont {Maier}}, \ and\ \bibinfo {author}
  {\bibfnamefont {M.~S.}\ \bibnamefont {Kim}},\ }\href {\doibase
  10.1038/nphys2615} {\bibfield  {journal} {\bibinfo  {journal} {Nature
  Physics}\ }\textbf {\bibinfo {volume} {9}},\ \bibinfo {pages} {329} (\bibinfo
  {year} {2013})}\BibitemShut {NoStop}%
\bibitem [{\citenamefont {Veselago}(1967)}]{Veselago1967}%
  \BibitemOpen
  \bibfield  {author} {\bibinfo {author} {\bibfnamefont {V.~G.}\ \bibnamefont
  {Veselago}},\ }\href {\doibase 10.3367/UFNr.0092.196707d.0517} {\bibfield
  {journal} {\bibinfo  {journal} {Uspekhi Fizicheskih Nauk}\ }\textbf {\bibinfo
  {volume} {92}},\ \bibinfo {pages} {517} (\bibinfo {year} {1967})}\BibitemShut
  {NoStop}%
\bibitem [{\citenamefont {Pendry}(2000)}]{PhysRevLett.85.3966}%
  \BibitemOpen
  \bibfield  {author} {\bibinfo {author} {\bibfnamefont {J.~B.}\ \bibnamefont
  {Pendry}},\ }\href {\doibase 10.1103/PhysRevLett.85.3966} {\bibfield
  {journal} {\bibinfo  {journal} {Phys. Rev. Lett.}\ }\textbf {\bibinfo
  {volume} {85}},\ \bibinfo {pages} {3966} (\bibinfo {year}
  {2000})}\BibitemShut {NoStop}%
\bibitem [{\citenamefont {Kalusniak}\ \emph {et~al.}(2014)\citenamefont
  {Kalusniak}, \citenamefont {Sadofev},\ and\ \citenamefont
  {Henneberger}}]{PhysRevLett.112.137401}%
  \BibitemOpen
  \bibfield  {author} {\bibinfo {author} {\bibfnamefont {S.}~\bibnamefont
  {Kalusniak}}, \bibinfo {author} {\bibfnamefont {S.}~\bibnamefont {Sadofev}},
  \ and\ \bibinfo {author} {\bibfnamefont {F.}~\bibnamefont {Henneberger}},\
  }\href {\doibase 10.1103/PhysRevLett.112.137401} {\bibfield  {journal}
  {\bibinfo  {journal} {Phys. Rev. Lett.}\ }\textbf {\bibinfo {volume} {112}},\
  \bibinfo {pages} {137401} (\bibinfo {year} {2014})}\BibitemShut {NoStop}%
\bibitem [{\citenamefont {Rayleigh}\ and\ \citenamefont
  {Strutt}(1897)}]{Rayleigh1897}%
  \BibitemOpen
  \bibfield  {author} {\bibinfo {author} {\bibnamefont {Rayleigh}}\ and\
  \bibinfo {author} {\bibfnamefont {J.~W.}\ \bibnamefont {Strutt}},\
  }\href@noop {} {\bibfield  {journal} {\bibinfo  {journal} {Phil. Mag.}\
  }\textbf {\bibinfo {volume} {XLIII}},\ \bibinfo {pages} {125} (\bibinfo
  {year} {1897})}\BibitemShut {NoStop}%
\bibitem [{\citenamefont {Sommerfeld}(1909)}]{Sommerfeld1909}%
  \BibitemOpen
  \bibfield  {author} {\bibinfo {author} {\bibfnamefont {A.}~\bibnamefont
  {Sommerfeld}},\ }\href@noop {} {\bibfield  {journal} {\bibinfo  {journal}
  {Annalen der Physik (4th series)}\ }\textbf {\bibinfo {volume} {28}},\
  \bibinfo {pages} {44} (\bibinfo {year} {1909})}\BibitemShut {NoStop}%
\bibitem [{\citenamefont {Stratton}(1941)}]{Stratton1941}%
  \BibitemOpen
  \bibfield  {author} {\bibinfo {author} {\bibfnamefont {I.~A.}\ \bibnamefont
  {Stratton}},\ }\href@noop {} {\emph {\bibinfo {title} {Electromagnetic
  Theory}}}\ (\bibinfo  {publisher} {McGraw-Hill},\ \bibinfo {year}
  {1941})\BibitemShut {NoStop}%
\bibitem [{\citenamefont {Kogelnik}(1988)}]{Kogelnik1988}%
  \BibitemOpen
  \bibfield  {author} {\bibinfo {author} {\bibfnamefont {H.}~\bibnamefont
  {Kogelnik}},\ }in\ \href@noop {} {\emph {\bibinfo {booktitle} {Guided-Wave
  Optoelectronics}}},\ \bibinfo {editor} {edited by\ \bibinfo {editor}
  {\bibfnamefont {T.}~\bibnamefont {Tamir}}}\ (\bibinfo  {publisher}
  {Springer-Verlag},\ \bibinfo {year} {1988})\ pp.\ \bibinfo {pages}
  {7--88}\BibitemShut {NoStop}%
\bibitem [{\citenamefont {Yeh}\ and\ \citenamefont
  {Shimabukuro}(2008)}]{Yeh2008}%
  \BibitemOpen
  \bibfield  {author} {\bibinfo {author} {\bibfnamefont {C.}~\bibnamefont
  {Yeh}}\ and\ \bibinfo {author} {\bibfnamefont {F.}~\bibnamefont
  {Shimabukuro}},\ }\href@noop {} {\emph {\bibinfo {title} {The Essence of
  Dielectric Waveguides}}}\ (\bibinfo  {publisher} {Springer Science+Business
  Media},\ \bibinfo {year} {2008})\BibitemShut {NoStop}%
\bibitem [{\citenamefont {Packard}(1984)}]{1132809}%
  \BibitemOpen
  \bibfield  {author} {\bibinfo {author} {\bibfnamefont {K.}~\bibnamefont
  {Packard}},\ }\href {\doibase 10.1109/TMTT.1984.1132809} {\bibfield
  {journal} {\bibinfo  {journal} {Microwave Theory and Techniques, IEEE
  Transactions on}\ }\textbf {\bibinfo {volume} {32}},\ \bibinfo {pages} {961}
  (\bibinfo {year} {1984})}\BibitemShut {NoStop}%
\bibitem [{\citenamefont {Buckley}\ and\ \citenamefont
  {Berini}(2009{\natexlab{a}})}]{Buckley2009}%
  \BibitemOpen
  \bibfield  {author} {\bibinfo {author} {\bibfnamefont {R.}~\bibnamefont
  {Buckley}}\ and\ \bibinfo {author} {\bibfnamefont {P.}~\bibnamefont
  {Berini}},\ }\href {\doibase 10.1109/JLT.2009.2017033} {\bibfield  {journal}
  {\bibinfo  {journal} {J. Light. Technol.}\ }\textbf {\bibinfo {volume}
  {27}},\ \bibinfo {pages} {2800} (\bibinfo {year}
  {2009}{\natexlab{a}})}\BibitemShut {NoStop}%
\bibitem [{\citenamefont {Bahadori}\ \emph {et~al.}(2014)\citenamefont
  {Bahadori}, \citenamefont {Eshaghian},\ and\ \citenamefont
  {Mehrany}}]{Bahadori2014}%
  \BibitemOpen
  \bibfield  {author} {\bibinfo {author} {\bibfnamefont {M.}~\bibnamefont
  {Bahadori}}, \bibinfo {author} {\bibfnamefont {A.}~\bibnamefont {Eshaghian}},
  \ and\ \bibinfo {author} {\bibfnamefont {K.}~\bibnamefont {Mehrany}},\ }\href
  {\doibase 10.1109/JLT.2014.2331325} {\bibfield  {journal} {\bibinfo
  {journal} {J. Light. Technol.}\ }\textbf {\bibinfo {volume} {32}},\ \bibinfo
  {pages} {2659} (\bibinfo {year} {2014})}\BibitemShut {NoStop}%
\bibitem [{\citenamefont {Zheng}\ \emph {et~al.}(2009)\citenamefont {Zheng},
  \citenamefont {Shi}, \citenamefont {Wang},\ and\ \citenamefont
  {Li}}]{Zheng2009}%
  \BibitemOpen
  \bibfield  {author} {\bibinfo {author} {\bibfnamefont {G.-g.}\ \bibnamefont
  {Zheng}}, \bibinfo {author} {\bibfnamefont {L.-x.}\ \bibnamefont {Shi}},
  \bibinfo {author} {\bibfnamefont {H.-l.}\ \bibnamefont {Wang}}, \ and\
  \bibinfo {author} {\bibfnamefont {X.-y.}\ \bibnamefont {Li}},\ }\href
  {\doibase 10.1016/j.optcom.2009.06.002} {\bibfield  {journal} {\bibinfo
  {journal} {Opt. Commun.}\ }\textbf {\bibinfo {volume} {282}},\ \bibinfo
  {pages} {4146} (\bibinfo {year} {2009})}\BibitemShut {NoStop}%
\bibitem [{\citenamefont {Berm{\'{u}}dez-Ure{\~{n}}a}\ \emph
  {et~al.}(2015)\citenamefont {Berm{\'{u}}dez-Ure{\~{n}}a}, \citenamefont
  {Gonzalez-Ballestero}, \citenamefont {Geiselmann}, \citenamefont {Marty},
  \citenamefont {Radko}, \citenamefont {Holmgaard}, \citenamefont {Alaverdyan},
  \citenamefont {Moreno}, \citenamefont {Garc{\'{\i}}a-Vidal}, \citenamefont
  {Bozhevolnyi},\ and\ \citenamefont {Quidant}}]{Bermudez-Urena2015}%
  \BibitemOpen
  \bibfield  {author} {\bibinfo {author} {\bibfnamefont {E.}~\bibnamefont
  {Berm{\'{u}}dez-Ure{\~{n}}a}}, \bibinfo {author} {\bibfnamefont
  {C.}~\bibnamefont {Gonzalez-Ballestero}}, \bibinfo {author} {\bibfnamefont
  {M.}~\bibnamefont {Geiselmann}}, \bibinfo {author} {\bibfnamefont
  {R.}~\bibnamefont {Marty}}, \bibinfo {author} {\bibfnamefont {I.~P.}\
  \bibnamefont {Radko}}, \bibinfo {author} {\bibfnamefont {T.}~\bibnamefont
  {Holmgaard}}, \bibinfo {author} {\bibfnamefont {Y.}~\bibnamefont
  {Alaverdyan}}, \bibinfo {author} {\bibfnamefont {E.}~\bibnamefont {Moreno}},
  \bibinfo {author} {\bibfnamefont {F.~J.}\ \bibnamefont
  {Garc{\'{\i}}a-Vidal}}, \bibinfo {author} {\bibfnamefont {S.~I.}\
  \bibnamefont {Bozhevolnyi}}, \ and\ \bibinfo {author} {\bibfnamefont
  {R.}~\bibnamefont {Quidant}},\ }\href {\doibase 10.1038/ncomms8883}
  {\bibfield  {journal} {\bibinfo  {journal} {Nat. Commun.}\ }\textbf {\bibinfo
  {volume} {6}},\ \bibinfo {pages} {7883} (\bibinfo {year} {2015})}\BibitemShut
  {NoStop}%
\bibitem [{\citenamefont {Berini}(2009)}]{Berini2009}%
  \BibitemOpen
  \bibfield  {author} {\bibinfo {author} {\bibfnamefont {P.}~\bibnamefont
  {Berini}},\ }\href {\doibase 10.1364/AOP.1.000484} {\bibfield  {journal}
  {\bibinfo  {journal} {Adv. Opt. Photonics}\ }\textbf {\bibinfo {volume}
  {1}},\ \bibinfo {pages} {484} (\bibinfo {year} {2009})}\BibitemShut {NoStop}%
\bibitem [{\citenamefont {Kurokawa}\ and\ \citenamefont
  {Miyazaki}(2007)}]{Kurokawa2007}%
  \BibitemOpen
  \bibfield  {author} {\bibinfo {author} {\bibfnamefont {Y.}~\bibnamefont
  {Kurokawa}}\ and\ \bibinfo {author} {\bibfnamefont {H.~T.}\ \bibnamefont
  {Miyazaki}},\ }\href {\doibase 10.1103/PhysRevB.75.035411} {\bibfield
  {journal} {\bibinfo  {journal} {Phys. Rev. B - Condens. Matter Mater. Phys.}\
  }\textbf {\bibinfo {volume} {75}},\ \bibinfo {pages} {1} (\bibinfo {year}
  {2007})}\BibitemShut {NoStop}%
\bibitem [{\citenamefont {Buckley}\ and\ \citenamefont
  {Berini}(2009{\natexlab{b}})}]{Buckley2009a}%
  \BibitemOpen
  \bibfield  {author} {\bibinfo {author} {\bibfnamefont {R.}~\bibnamefont
  {Buckley}}\ and\ \bibinfo {author} {\bibfnamefont {P.}~\bibnamefont
  {Berini}},\ }\href {\doibase 10.1364/OL.34.000223} {\bibfield  {journal}
  {\bibinfo  {journal} {Opt. Lett.}\ }\textbf {\bibinfo {volume} {34}},\
  \bibinfo {pages} {223} (\bibinfo {year} {2009}{\natexlab{b}})}\BibitemShut
  {NoStop}%
\bibitem [{\citenamefont {Lavoie}\ \emph {et~al.}(2013)\citenamefont {Lavoie},
  \citenamefont {Leung},\ and\ \citenamefont {Sanders}}]{pra-2013}%
  \BibitemOpen
  \bibfield  {author} {\bibinfo {author} {\bibfnamefont {B.~R.}\ \bibnamefont
  {Lavoie}}, \bibinfo {author} {\bibfnamefont {P.~M.}\ \bibnamefont {Leung}}, \
  and\ \bibinfo {author} {\bibfnamefont {B.~C.}\ \bibnamefont {Sanders}},\
  }\href@noop {} {\bibfield  {journal} {\bibinfo  {journal} {Physical Review
  A}\ }\textbf {\bibinfo {volume} {88}},\ \bibinfo {pages} {023860} (\bibinfo
  {year} {2013})}\BibitemShut {NoStop}%
\bibitem [{\citenamefont {Lavoie}\ \emph {et~al.}(2012)\citenamefont {Lavoie},
  \citenamefont {Leung},\ and\ \citenamefont {Sanders}}]{Lavoie2012}%
  \BibitemOpen
  \bibfield  {author} {\bibinfo {author} {\bibfnamefont {B.~R.}\ \bibnamefont
  {Lavoie}}, \bibinfo {author} {\bibfnamefont {P.~M.}\ \bibnamefont {Leung}}, \
  and\ \bibinfo {author} {\bibfnamefont {B.~C.}\ \bibnamefont {Sanders}},\
  }\href {\doibase http://dx.doi.org/10.1016/j.photonics.2012.05.010}
  {\bibfield  {journal} {\bibinfo  {journal} {Photonics and Nanostructures -
  Fundamentals and Applications}\ }\textbf {\bibinfo {volume} {10}},\ \bibinfo
  {pages} {602 } (\bibinfo {year} {2012})},\ \bibinfo {note} {taCoNa-Photonics
  2011}\BibitemShut {NoStop}%
\bibitem [{\citenamefont {Ashour}(2013)}]{Ashour2013}%
  \BibitemOpen
  \bibfield  {author} {\bibinfo {author} {\bibfnamefont {H.~S.}\ \bibnamefont
  {Ashour}},\ }\href {\doibase 10.4236/jmp.2013.49156} {\bibfield  {journal}
  {\bibinfo  {journal} {J. Mod. Phys.}\ }\textbf {\bibinfo {volume} {04}},\
  \bibinfo {pages} {1165} (\bibinfo {year} {2013})}\BibitemShut {NoStop}%
\bibitem [{\citenamefont {Han}\ and\ \citenamefont
  {Bozhevolnyi}(2014)}]{Han2014}%
  \BibitemOpen
  \bibfield  {author} {\bibinfo {author} {\bibfnamefont {Z.}~\bibnamefont
  {Han}}\ and\ \bibinfo {author} {\bibfnamefont {S.~I.}\ \bibnamefont
  {Bozhevolnyi}},\ }\href {\doibase 10.1016/B978-0-444-59526-3.00005-7} {\emph
  {\bibinfo {title} {Mod. Plasmon.}}},\ Vol.~\bibinfo {volume} {4}\ (\bibinfo
  {publisher} {Elsevier B.V.},\ \bibinfo {year} {2014})\ Chap.~\bibinfo
  {chapter} {5}, pp.\ \bibinfo {pages} {137--187}\BibitemShut {NoStop}%
\bibitem [{\citenamefont {Chen}\ \emph {et~al.}(2015)\citenamefont {Chen},
  \citenamefont {Ke}, \citenamefont {Lan},\ and\ \citenamefont
  {Chan}}]{Chen2015}%
  \BibitemOpen
  \bibfield  {author} {\bibinfo {author} {\bibfnamefont {C.-M.}\ \bibnamefont
  {Chen}}, \bibinfo {author} {\bibfnamefont {J.-L.}\ \bibnamefont {Ke}},
  \bibinfo {author} {\bibfnamefont {Y.-C.}\ \bibnamefont {Lan}}, \ and\
  \bibinfo {author} {\bibfnamefont {M.-C.}\ \bibnamefont {Chan}},\ }\href
  {\doibase 10.1364/OE.23.029321} {\bibfield  {journal} {\bibinfo  {journal}
  {Opt. Express}\ }\textbf {\bibinfo {volume} {23}},\ \bibinfo {pages} {29321}
  (\bibinfo {year} {2015})}\BibitemShut {NoStop}%
\bibitem [{\citenamefont {Yang}\ \emph
  {et~al.}(2015{\natexlab{a}})\citenamefont {Yang}, \citenamefont {Li},\ and\
  \citenamefont {Xiao}}]{Yang2015}%
  \BibitemOpen
  \bibfield  {author} {\bibinfo {author} {\bibfnamefont {H.}~\bibnamefont
  {Yang}}, \bibinfo {author} {\bibfnamefont {J.}~\bibnamefont {Li}}, \ and\
  \bibinfo {author} {\bibfnamefont {G.}~\bibnamefont {Xiao}},\ }\href {\doibase
  10.1016/j.optcom.2015.10.057} {\bibfield  {journal} {\bibinfo  {journal}
  {Opt. Commun.}\ } (\bibinfo {year} {2015}{\natexlab{a}}),\
  10.1016/j.optcom.2015.10.057}\BibitemShut {NoStop}%
\bibitem [{\citenamefont {{De Broglie}}(1941)}]{DeBroglie1941}%
  \BibitemOpen
  \bibfield  {author} {\bibinfo {author} {\bibfnamefont {L.}~\bibnamefont {{De
  Broglie}}},\ }\href@noop {} {\emph {\bibinfo {title} {{Probl\`{e}mes de
  propagations guid\'{e}es des ondes \'{e}lectromagn\'{e}tiques}}}}\ (\bibinfo
  {publisher} {Gauthier-Villars},\ \bibinfo {address} {Paris},\ \bibinfo {year}
  {1941})\ p.\ \bibinfo {pages} {114}\BibitemShut {NoStop}%
\bibitem [{\citenamefont {Rivlin}(1997)}]{Rivlin1997}%
  \BibitemOpen
  \bibfield  {author} {\bibinfo {author} {\bibfnamefont {L.~A.}\ \bibnamefont
  {Rivlin}},\ }\href {\doibase 10.3367/UFNr.0167.199703g.0309} {\bibfield
  {journal} {\bibinfo  {journal} {Uspekhi Fizicheskih Nauk}\ }\textbf {\bibinfo
  {volume} {167}},\ \bibinfo {pages} {309} (\bibinfo {year}
  {1997})}\BibitemShut {NoStop}%
\bibitem [{\citenamefont {Sambles}(2015)}]{Sambles2015}%
  \BibitemOpen
  \bibfield  {author} {\bibinfo {author} {\bibfnamefont {J.~R.}\ \bibnamefont
  {Sambles}},\ }\href {\doibase 10.1126/science.aaa6931} {\bibfield  {journal}
  {\bibinfo  {journal} {Science}\ }\textbf {\bibinfo {volume} {347}},\ \bibinfo
  {pages} {828} (\bibinfo {year} {2015})}\BibitemShut {NoStop}%
\bibitem [{\citenamefont {Klaers}\ \emph {et~al.}(2010)\citenamefont {Klaers},
  \citenamefont {Schmitt}, \citenamefont {Vewinger},\ and\ \citenamefont
  {Weitz}}]{Klaers2010}%
  \BibitemOpen
  \bibfield  {author} {\bibinfo {author} {\bibfnamefont {J.}~\bibnamefont
  {Klaers}}, \bibinfo {author} {\bibfnamefont {J.}~\bibnamefont {Schmitt}},
  \bibinfo {author} {\bibfnamefont {F.}~\bibnamefont {Vewinger}}, \ and\
  \bibinfo {author} {\bibfnamefont {M.}~\bibnamefont {Weitz}},\ }\href
  {\doibase 10.1038/nature09567} {\bibfield  {journal} {\bibinfo  {journal}
  {Nature}\ }\textbf {\bibinfo {volume} {468}},\ \bibinfo {pages} {545}
  (\bibinfo {year} {2010})},\ \Eprint {http://arxiv.org/abs/1007.4088}
  {1007.4088} \BibitemShut {NoStop}%
\bibitem [{\citenamefont {Kacenelenbaum}(1953)}]{Kacenelenbaum:1953}%
  \BibitemOpen
  \bibfield  {author} {\bibinfo {author} {\bibfnamefont {B.}~\bibnamefont
  {Kacenelenbaum}},\ }\href@noop {} {\bibfield  {journal} {\bibinfo  {journal}
  {Doklady AN USSR (In Russian)}\ }\textbf {\bibinfo {volume} {88}},\ \bibinfo
  {pages} {37} (\bibinfo {year} {1953})}\BibitemShut {NoStop}%
\bibitem [{\citenamefont {Born}\ and\ \citenamefont
  {Wolf}(1999)}]{Born-Wolf:1999:PO}%
  \BibitemOpen
  \bibfield  {author} {\bibinfo {author} {\bibfnamefont {M.}~\bibnamefont
  {Born}}\ and\ \bibinfo {author} {\bibfnamefont {E.}~\bibnamefont {Wolf}},\
  }\href@noop {} {\emph {\bibinfo {title} {Principles of Optics}}}\ (\bibinfo
  {publisher} {Cambridge University Press},\ \bibinfo {year}
  {1999})\BibitemShut {NoStop}%
\bibitem [{\citenamefont {Adams}(1981)}]{Adams1981}%
  \BibitemOpen
  \bibfield  {author} {\bibinfo {author} {\bibfnamefont {M.~J.}\ \bibnamefont
  {Adams}},\ }\href@noop {} {\emph {\bibinfo {title} {An Introduction to
  optical waveguides}}}\ (\bibinfo  {publisher} {Wiley},\ \bibinfo {year}
  {1981})\BibitemShut {NoStop}%
\bibitem [{\citenamefont {Yuferev}\ and\ \citenamefont
  {Ida}(2008)}]{Yuferev:2008:SIBc}%
  \BibitemOpen
  \bibfield  {author} {\bibinfo {author} {\bibfnamefont {S.}~\bibnamefont
  {Yuferev}}\ and\ \bibinfo {author} {\bibfnamefont {N.}~\bibnamefont {Ida}},\
  }\href@noop {} {\emph {\bibinfo {title} {Surface impedance boundary
  conditions (A Comprehensive Approach)}}}\ (\bibinfo  {publisher} {CRC
  Press},\ \bibinfo {year} {2008})\BibitemShut {NoStop}%
\bibitem [{\citenamefont {Novotny}\ and\ \citenamefont
  {Hafner}(1994)}]{Novotny1994}%
  \BibitemOpen
  \bibfield  {author} {\bibinfo {author} {\bibfnamefont {L.}~\bibnamefont
  {Novotny}}\ and\ \bibinfo {author} {\bibfnamefont {C.}~\bibnamefont
  {Hafner}},\ }\href {\doibase 10.1103/PhysRevE.50.4094} {\bibfield  {journal}
  {\bibinfo  {journal} {Phys. Rev. E}\ }\textbf {\bibinfo {volume} {50}},\
  \bibinfo {pages} {4094} (\bibinfo {year} {1994})}\BibitemShut {NoStop}%
\bibitem [{\citenamefont {Kim}\ \emph {et~al.}(2014)\citenamefont {Kim},
  \citenamefont {Wong},\ and\ \citenamefont
  {Eleftheriades}}]{PhysRevX.4.041042}%
  \BibitemOpen
  \bibfield  {author} {\bibinfo {author} {\bibfnamefont {M.}~\bibnamefont
  {Kim}}, \bibinfo {author} {\bibfnamefont {A.~M.~H.}\ \bibnamefont {Wong}}, \
  and\ \bibinfo {author} {\bibfnamefont {G.~V.}\ \bibnamefont
  {Eleftheriades}},\ }\href {\doibase 10.1103/PhysRevX.4.041042} {\bibfield
  {journal} {\bibinfo  {journal} {Phys. Rev. X}\ }\textbf {\bibinfo {volume}
  {4}},\ \bibinfo {pages} {041042} (\bibinfo {year} {2014})}\BibitemShut
  {NoStop}%
\bibitem [{\citenamefont {Kamli}\ \emph {et~al.}(2008)\citenamefont {Kamli},
  \citenamefont {Moiseev},\ and\ \citenamefont {Sanders}}]{Kamli2008}%
  \BibitemOpen
  \bibfield  {author} {\bibinfo {author} {\bibfnamefont {A.}~\bibnamefont
  {Kamli}}, \bibinfo {author} {\bibfnamefont {S.~A.}\ \bibnamefont {Moiseev}},
  \ and\ \bibinfo {author} {\bibfnamefont {B.~C.}\ \bibnamefont {Sanders}},\
  }\href {\doibase 10.1103/PhysRevLett.101.263601} {\bibfield  {journal}
  {\bibinfo  {journal} {Phys. Rev. Lett.}\ }\textbf {\bibinfo {volume} {101}}
  (\bibinfo {year} {2008}),\ 10.1103/PhysRevLett.101.263601}\BibitemShut
  {NoStop}%
\bibitem [{\citenamefont {Moiseev}\ \emph {et~al.}(2010)\citenamefont
  {Moiseev}, \citenamefont {Kamli},\ and\ \citenamefont
  {Sanders}}]{PhysRevA.81.033839}%
  \BibitemOpen
  \bibfield  {author} {\bibinfo {author} {\bibfnamefont {S.~A.}\ \bibnamefont
  {Moiseev}}, \bibinfo {author} {\bibfnamefont {A.~A.}\ \bibnamefont {Kamli}},
  \ and\ \bibinfo {author} {\bibfnamefont {B.~C.}\ \bibnamefont {Sanders}},\
  }\href {\doibase 10.1103/PhysRevA.81.033839} {\bibfield  {journal} {\bibinfo
  {journal} {Phys. Rev. A}\ }\textbf {\bibinfo {volume} {81}},\ \bibinfo
  {pages} {033839} (\bibinfo {year} {2010})}\BibitemShut {NoStop}%
\bibitem [{\citenamefont {Guglielmi}(2010)}]{Guglielmi2010}%
  \BibitemOpen
  \bibfield  {author} {\bibinfo {author} {\bibfnamefont {A.}~\bibnamefont
  {Guglielmi}},\ }\href {\doibase 10.3367/UFNr.0180.201001g.0105} {\bibfield
  {journal} {\bibinfo  {journal} {Uspekhi Fiz. Nauk}\ }\textbf {\bibinfo
  {volume} {180}},\ \bibinfo {pages} {105} (\bibinfo {year}
  {2010})}\BibitemShut {NoStop}%
\bibitem [{\citenamefont {Arslanov}(2006)}]{Arslanov2005}%
  \BibitemOpen
  \bibfield  {author} {\bibinfo {author} {\bibfnamefont {N.~M.}\ \bibnamefont
  {Arslanov}},\ }\href {\doibase 10.1088/1464-4258/8/3/018} {\bibfield
  {journal} {\bibinfo  {journal} {J. Opt. A Pure Appl. Opt.}\ }\textbf
  {\bibinfo {volume} {8}},\ \bibinfo {pages} {338} (\bibinfo {year} {2006})},\
  \Eprint {http://arxiv.org/abs/0509209} {arXiv:0509209 [physics]} \BibitemShut
  {NoStop}%
\bibitem [{\citenamefont {Arslanov}\ and\ \citenamefont
  {Moiseev}(2007)}]{Arslanov2007}%
  \BibitemOpen
  \bibfield  {author} {\bibinfo {author} {\bibfnamefont {N.~M.}\ \bibnamefont
  {Arslanov}}\ and\ \bibinfo {author} {\bibfnamefont {S.~A.}\ \bibnamefont
  {Moiseev}},\ }\href {\doibase 10.1364/JOSAA.24.000831} {\bibfield  {journal}
  {\bibinfo  {journal} {J. Opt. Soc. Am. A. Opt. Image Sci. Vis.}\ }\textbf
  {\bibinfo {volume} {24}},\ \bibinfo {pages} {831} (\bibinfo {year} {2007})},\
  \Eprint {http://arxiv.org/abs/0509187} {arXiv:0509187 [physics]} \BibitemShut
  {NoStop}%
\bibitem [{\citenamefont {Fok}(1948)}]{Fok1948}%
  \BibitemOpen
  \bibfield  {author} {\bibinfo {author} {\bibfnamefont {V.}~\bibnamefont
  {Fok}},\ }\href {\doibase 10.3367/UFNr.0036.194811d.0308} {\bibfield
  {journal} {\bibinfo  {journal} {Uspekhi Fiz. Nauk}\ }\textbf {\bibinfo
  {volume} {36}},\ \bibinfo {pages} {308} (\bibinfo {year} {1948})}\BibitemShut
  {NoStop}%
\bibitem [{\citenamefont {Sun}\ \emph {et~al.}(2011)\citenamefont {Sun},
  \citenamefont {Videen}, \citenamefont {Lin}, \citenamefont {Hu},\ and\
  \citenamefont {Fu}}]{Sun2011}%
  \BibitemOpen
  \bibfield  {author} {\bibinfo {author} {\bibfnamefont {W.}~\bibnamefont
  {Sun}}, \bibinfo {author} {\bibfnamefont {G.}~\bibnamefont {Videen}},
  \bibinfo {author} {\bibfnamefont {B.}~\bibnamefont {Lin}}, \bibinfo {author}
  {\bibfnamefont {Y.}~\bibnamefont {Hu}}, \ and\ \bibinfo {author}
  {\bibfnamefont {Q.}~\bibnamefont {Fu}},\ }\href {\doibase
  10.1016/j.jqsrt.2010.03.009} {\bibfield  {journal} {\bibinfo  {journal} {J.
  Quant. Spectrosc. Radiat. Transf.}\ }\textbf {\bibinfo {volume} {112}},\
  \bibinfo {pages} {174} (\bibinfo {year} {2011})}\BibitemShut {NoStop}%
\bibitem [{\citenamefont {Chang}\ \emph {et~al.}(2006)\citenamefont {Chang},
  \citenamefont {S{\o}rensen}, \citenamefont {Hemmer},\ and\ \citenamefont
  {Lukin}}]{Chang2006}%
  \BibitemOpen
  \bibfield  {author} {\bibinfo {author} {\bibfnamefont {D.~E.}\ \bibnamefont
  {Chang}}, \bibinfo {author} {\bibfnamefont {A.~S.}\ \bibnamefont
  {S{\o}rensen}}, \bibinfo {author} {\bibfnamefont {P.~R.}\ \bibnamefont
  {Hemmer}}, \ and\ \bibinfo {author} {\bibfnamefont {M.~D.}\ \bibnamefont
  {Lukin}},\ }\href {\doibase 10.1103/PhysRevLett.97.053002} {\bibfield
  {journal} {\bibinfo  {journal} {Phys. Rev. Lett.}\ }\textbf {\bibinfo
  {volume} {97}},\ \bibinfo {pages} {1} (\bibinfo {year} {2006})}\BibitemShut
  {NoStop}%
\bibitem [{\citenamefont {Nieddu}\ \emph {et~al.}(2015)\citenamefont {Nieddu},
  \citenamefont {Gokhroo},\ and\ \citenamefont {Chormaic}}]{Nieddu2015}%
  \BibitemOpen
  \bibfield  {author} {\bibinfo {author} {\bibfnamefont {T.}~\bibnamefont
  {Nieddu}}, \bibinfo {author} {\bibfnamefont {V.}~\bibnamefont {Gokhroo}}, \
  and\ \bibinfo {author} {\bibfnamefont {S.~N.}\ \bibnamefont {Chormaic}},\
  }\href {http://arxiv.org/pdf/1512.02753v1.pdf
  http://arxiv.org/abs/1512.02753} {\  (\bibinfo {year} {2015})},\ \Eprint
  {http://arxiv.org/abs/1512.02753} {arXiv:1512.02753} \BibitemShut {NoStop}%
\bibitem [{\citenamefont {Cardano}\ and\ \citenamefont
  {Marrucci}(2015)}]{Cardano2015}%
  \BibitemOpen
  \bibfield  {author} {\bibinfo {author} {\bibfnamefont {F.}~\bibnamefont
  {Cardano}}\ and\ \bibinfo {author} {\bibfnamefont {L.}~\bibnamefont
  {Marrucci}},\ }\href {\doibase 10.1038/nphoton.2015.232} {\bibfield
  {journal} {\bibinfo  {journal} {Nat. Photonics}\ }\textbf {\bibinfo {volume}
  {9}},\ \bibinfo {pages} {776} (\bibinfo {year} {2015})}\BibitemShut {NoStop}%
\bibitem [{\citenamefont {Bliokh}\ \emph {et~al.}(2015)\citenamefont {Bliokh},
  \citenamefont {Rodr{\'{\i}}guez-Fortu{\~{n}}o}, \citenamefont {Nori},\ and\
  \citenamefont {Zayats}}]{Bliokh2015}%
  \BibitemOpen
  \bibfield  {author} {\bibinfo {author} {\bibfnamefont {K.~Y.}\ \bibnamefont
  {Bliokh}}, \bibinfo {author} {\bibfnamefont {F.~J.}\ \bibnamefont
  {Rodr{\'{\i}}guez-Fortu{\~{n}}o}}, \bibinfo {author} {\bibfnamefont
  {F.}~\bibnamefont {Nori}}, \ and\ \bibinfo {author} {\bibfnamefont {A.~V.}\
  \bibnamefont {Zayats}},\ }\href {\doibase 10.1038/nphoton.2015.201}
  {\bibfield  {journal} {\bibinfo  {journal} {Nat. Photonics}\ }\textbf
  {\bibinfo {volume} {9}},\ \bibinfo {pages} {796} (\bibinfo {year}
  {2015})}\BibitemShut {NoStop}%
\bibitem [{\citenamefont {Sukhov}\ \emph {et~al.}(2015)\citenamefont {Sukhov},
  \citenamefont {Kajorndejnukul}, \citenamefont {Naraghi},\ and\ \citenamefont
  {Dogariu}}]{Sukhov2015}%
  \BibitemOpen
  \bibfield  {author} {\bibinfo {author} {\bibfnamefont {S.}~\bibnamefont
  {Sukhov}}, \bibinfo {author} {\bibfnamefont {V.}~\bibnamefont
  {Kajorndejnukul}}, \bibinfo {author} {\bibfnamefont {R.~R.}\ \bibnamefont
  {Naraghi}}, \ and\ \bibinfo {author} {\bibfnamefont {A.}~\bibnamefont
  {Dogariu}},\ }\href {\doibase 10.1038/nphoton.2015.200} {\bibfield  {journal}
  {\bibinfo  {journal} {Nat. Photonics}\ }\textbf {\bibinfo {volume} {9}},\
  \bibinfo {pages} {809} (\bibinfo {year} {2015})}\BibitemShut {NoStop}%
\bibitem [{\citenamefont {Manassah}(2012)}]{Manassah2012}%
  \BibitemOpen
  \bibfield  {author} {\bibinfo {author} {\bibfnamefont {J.~T.}\ \bibnamefont
  {Manassah}},\ }\href {\doibase 10.1016/j.cplett.2012.07.010} {\bibfield
  {journal} {\bibinfo  {journal} {Chem. Phys. Lett.}\ }\textbf {\bibinfo
  {volume} {544}},\ \bibinfo {pages} {73} (\bibinfo {year} {2012})}\BibitemShut
  {NoStop}%
\bibitem [{\citenamefont {Yang}\ \emph
  {et~al.}(2015{\natexlab{b}})\citenamefont {Yang}, \citenamefont {An},
  \citenamefont {Zhang}, \citenamefont {Chen},\ and\ \citenamefont
  {Oh}}]{Yang2015B}%
  \BibitemOpen
  \bibfield  {author} {\bibinfo {author} {\bibfnamefont {W.-l.}\ \bibnamefont
  {Yang}}, \bibinfo {author} {\bibfnamefont {J.-H.}\ \bibnamefont {An}},
  \bibinfo {author} {\bibfnamefont {C.-j.}\ \bibnamefont {Zhang}}, \bibinfo
  {author} {\bibfnamefont {C.-y.}\ \bibnamefont {Chen}}, \ and\ \bibinfo
  {author} {\bibfnamefont {C.~H.}\ \bibnamefont {Oh}},\ }\href {\doibase
  10.1038/srep15513} {\bibfield  {journal} {\bibinfo  {journal} {Sci. Rep.}\
  }\textbf {\bibinfo {volume} {5}},\ \bibinfo {pages} {15513} (\bibinfo {year}
  {2015}{\natexlab{b}})}\BibitemShut {NoStop}%
\end{thebibliography}%

\end{document}